\documentclass[twocolumn,showpacs,preprintnumbers,amsmath,amssymb,prl]{revtex4}% PRL

\usepackage{graphicx}% Include figure files
\usepackage{dcolumn}% Align table columns on decimal point
\usepackage{bm}% bold math
\begin{document}

%\preprint{APS/123-QED}

\title{Formation of nematic liquid crystalline phase of F-actin varies from continuous to biphasic transition} % Force line breaks with \\

\author{Jorge Viamontes}

\author{Jay X. Tang} 
\email[Correspondence address:]{Jay_Tang@Brown.edu}
\affiliation{Deparment of Physics, Brown University, Rhode Island 02912, USA}

\date{\today}	% It is always \today, today,
             %  but any date may be explicitly specified

\begin{abstract}
We show that the isotropic to nematic liquid crystalline phase transition of F-actin can be either continuous or discontinuous, depending critically on the filament length. For F-actin with average filament length $\geq 3 \mu m$, we confirm that the transition is continuous in both filament alignment and local concentration. In contrast, for filament length $\leq 2 \mu m$ the F-actin solution undergoes a first order transition. Tactoidal droplets of co-existing isotropic and nematic domains were observed. Phenomena of nucleation-and-growth and spinodal decomposition both occur, depending sensitively on the exact concentration and average filament length of F-actin. In the late stage, the tactoidal droplets continually grow and occasionally coalesce to form larger granules. 

\end{abstract}

% PACS, the Physics and Astronomy
                             % Classification Scheme.
%\keywords{Suggested keywords}%Use showkeys class option if keyword
                              %display desired
\maketitle

Cytoskeletal protein actin is responsible for cell morphology and motility  \cite{Molecular_biology_of_the_cell}. Globular actin (G-actin) polymerizes to form long filaments, F-actin. F-actin has a diameter of 8 nm \cite{HolmesPoppGebhardKabsch} and a distribution of lengths characteristic of the stochastic polymerization process \cite{Sept_Xu_Pollard_McCammon}. F-actin is a semiflexible polymer with a persistence length of 15-18 $\mu m$ \cite{Isambert_Vernier_Maggs_Carlier,Gittes_Mickey_Nettlenton_Howard}, which is larger than their average length in cells or \textit{in vitro}. There have been extensive studies concerning many remarkable properties of F-actin, including dynamic filament assembly and dissembly \cite{Fujiwara2002}, phase transitions \cite{Coppin, Furakawa, Viamontes2003}, and rheology \cite{GardelPRL2003}. Many of these properties are shared by other self-assembled protein filaments such as microtubules and collagen-based intracellular matrix \cite{Molecular_biology_of_the_cell}.  

Of particular relevance to this report is that F-actin undergoes an isotropic (I) to nematic (N) liquid crystalline phase transition. The onset concentration of the transition is inversely proportional to the average filament length $\ell$ \cite{Coppin, Furakawa, Suzuki, Viamontes2003}, consistent with statistical mechanical theories \cite{Onsager,Flory}. The experimental studies use optical birefringence methods to measure the F-actin alignment across the I-N transition region. Under certain preparation conditions, "zebra" birefringence patterns were observed, which have been attributed to the spontaneous separation of F-actin into I and N domains \cite{Suzuki}. Two more recent studies show \cite{Coppin, Viamontes2003}, however, that the F-actin I-N transition appears to be continuous in both filament alignment and concentration for $\ell \geq 3 \mu m$. It has been argued that perhaps due to the extreme filament length, polydispersity, and semi flexibility, a combined outcome of defect suppression and entanglement renders the F-actin I-N transition continuous \cite{Viamontes2003}. The phenomenon may be relevant to the theory of Lammert, Rokshar, and Toner (LRT) \cite{LRTPRL,LRTPRE}, which predicts that a high disclination line defect energy renders the I-N transition into two continuous ones.

In this paper we confirm the continuous features of the I-N transition for solutions of long F-actin ($\ell \geq 3 \mu m$), but also show for the first time the I-N co-existence and domain separation of F-actin for solutions with $\ell \leq 2 \mu m$. Tactoidal droplets were observed in at least three types: N tactoids in an I background, I tactoids in an N background, coexistence of N and I tactoids on a uniform backgroud of weak alignment. The droplets grow in two distinct ways:  1. nucleation of scattered tactoids and growth; 2. spinodal decomposition, followed by coasening. In the late stage for both cases, coalescence of tactoidal droplets were observed and studied. Slow axis measurements of the N tactoids suggest that the director field smoothly follows the surface contour, connecting point defects at two opposite poles, which are called boojums  \cite{Prinsen_van_der_Schoot,Drzaic}

\begin{figure*}[t]
\includegraphics[width=17.0cm]{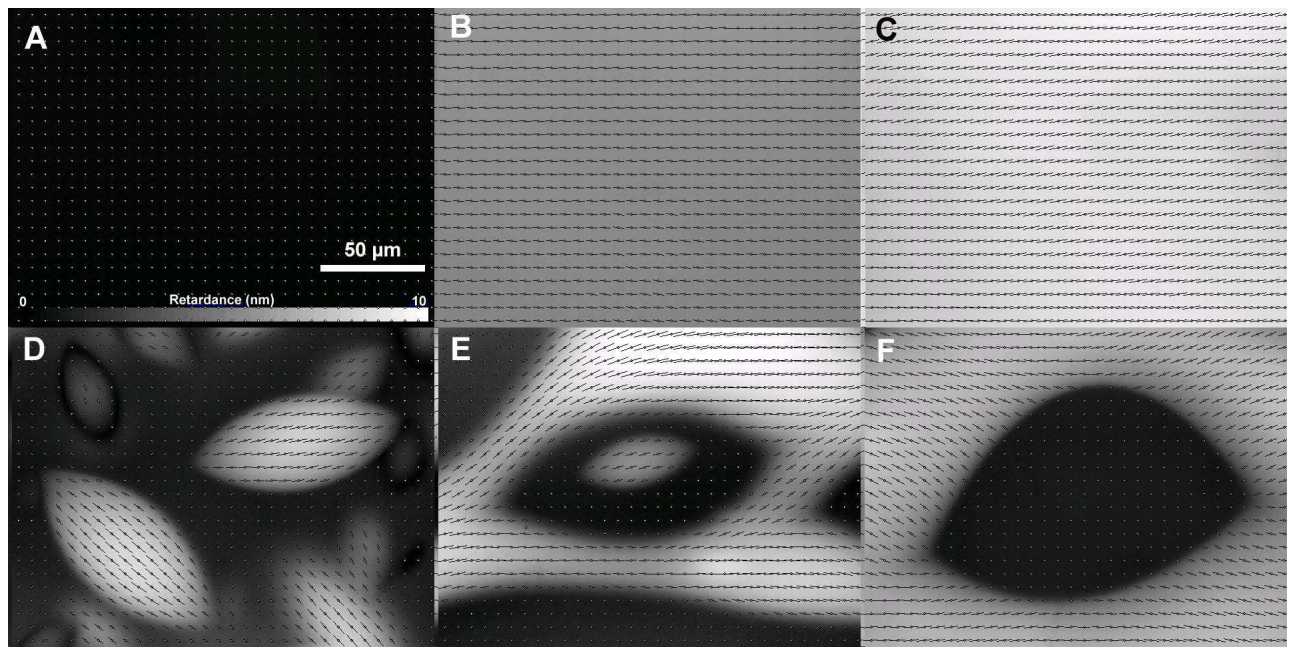}
\caption{\label{fig1}Birefringence measurement of F-actin solutions at $\ell=11  \mu m$ (\textbf{A,B,C}) and at $\ell=1  \mu m$ (\textbf{D,E,F}). Gray scale intensity bar indicates retardance values in the range of 0 to 10 nm. The line segments represent the local direction of filament alignment. (\textbf{A,B,C}) Representative samples at the I phase, in the I-N transition region for long F-actin, and at the N phase, respectively. No discontinuity is detected in the transition region. (\textbf{D,E,F}) At $\ell=1  \mu m$ and in the I-N transition region, F-actin phase separates into tactoidal droplets: N droplet in an I background (\textbf{D}), coexistence of N and I droplets (\textbf{E}), and an I  droplet in a N background (\textbf{F}).}
\end{figure*}

G-actin was extracted from rabbit skeletal muscle following an established method \cite{Pardee}. F-actin $\ell$ was varied by addition of gelsolin, a filament severing and end-capping protein \cite{JanmeyPeetermans,TangJanmey}. G-actin was polymerized upon addition of KCl and MgCl$_2$ upto 50 mM and 2 mM, respectively. Rectangular capillary tubes from VitroCom Inc. (Mt. Lks., N.J) of crossectional dimensions 0.2$\times$2 mm were used for measurements by fluorescence and birefringence microscopy. Both ends of the capillary tube were sealed by an inert glue to eliminate flow. Birefringence measurements were performed on a Nikon E-800 microscope equipped with the CRI PolScope package \cite{Viamontes2003}. PolScope is capable of measuring the optical birefringence and the direction of slow axis at each pixel position, thus reporting local alignment \cite{OldenbourgMei, ShribakOldenbourg}. F-actin was labeled 1 to 1000 with TRITC-Phalloidin (Sigma, St Louis, MO) for fluorescence measurements, performed as previously described \cite{Viamontes2003}. 2D Fast Fourier Transform (2D-FFT) analysis was performed similarly to Bees and Hill \cite{Bees_and_Hill}, using the MatLab 7.0 software (The MathWorks, Inc.).  

Different features are observed between samples of several $\ell$ of F-actin in their respective ranges of concentration over which the I-N transition occurs. F-actin with no gelsolin added were measured to be of $\ell=11  \mu m$. Fig.~\ref{fig1} shows representative results of birefringence and filament alignment of F-actin in the I phase (Fig.~\ref{fig1}\textbf{A}), transition region (Fig.~\ref{fig1}\textbf{B}), and the N phase (Fig.~\ref{fig1}\textbf{C}). Of particular note is that in the I-N transition region uniform retardance is found, suggesting that F-actin is continuous in alignment (Fig.~\ref{fig1}\textbf{B}) and in concentration \cite{Viamontes2003}. Zebra patterns are occationally observed, especially near the wall of a capillary tube, or at an air liquid interface, examples of which have been shown by previous studies \cite{Coppin, Suzuki, Viamontes2003}. Even at the location of a zebra pattern, the local concentration of actin remains uniform, suggesting a lack of co-existence in long F-actin samples \cite{Viamontes2003}. In contrast, upon further reduction of $\ell$ to $\leq 2 \mu m$ the F-actin solution phase separates into tactoidal droplets and their surrounding medium (Fig.~\ref{fig1}\textbf{D}). An increase in concentration gives rise to co-existence of I and N tactoids (Fig.~\ref{fig1}\textbf{E}), and I tactoids in N background (Fig.~\ref{fig1}\textbf{F}). Local concentration of F-actin is measured by quantitative fluorescence labelling, which  shows that N tactoids are denser in protein than the surrounding background, but only by 20-30 \%. The actin concentration within the I tactoids is lower than the sorrunding N region by a similar percentage. These results comfirm the weakly first order nature of the I-N transition.

\begin{figure}[t]
\includegraphics[width=8.0cm]{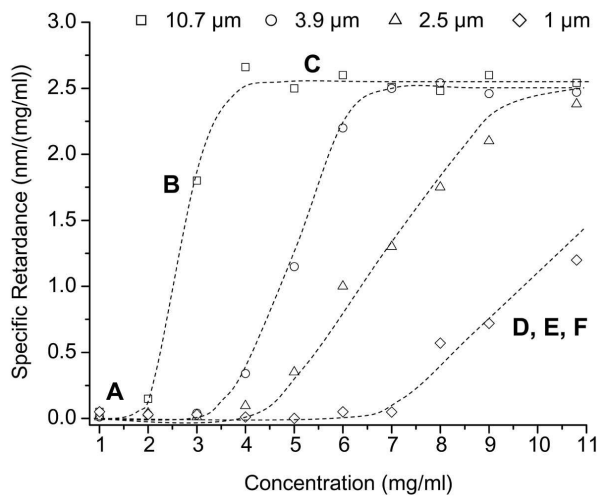}
\caption{\label{fig2}Specific retardance of F-actin as a function of concentration with varying $\ell$. $\Box$ represents F-actin solution with no added Gelsolin. $\bigcirc$, $\triangle$, and $\Diamond$ represent $\ell$ = 3.9, 2.5, and 1$\mu m$, respectively. The letter designations also correspond to regions noted in Fig.~\ref{fig1}. Dashed lines are guides to the eye.}
\end{figure}

We have measured the average alignment of F-actin at four $\ell$ as a function of actin concentration over the range of I-N transition (Fig.~\ref{fig2}). $\ell$ was determined by fluorescence imaging or AFM (for the shortest $\ell$) of single F-actin for at least 500 filaments for each $\ell$. Below a threshold concentration F-actin solution is in the I phase, thus the retardance is zero (Region \textbf{A}). As the  concentration increases, the solution reaches the I-N transition region, characterized by the sharp increase of specific retardance (Region \textbf{B}). In the high concentration region, F-actin solution is completely in the N phase (Region \textbf{C}). As $\ell$ decreases, the onset concentration of the I-N transition increases, consistent with the earlier reports \cite{Coppin, Furakawa, Suzuki, Viamontes2003}.  \textbf{D,E,F} on Fig. \ref{fig2} indicate the I-N transition region for $\ell = 1 \mu m$, where the specific birefringence values were measured prior to phase separation as shown in Fig. \ref{fig1}.

%polymerized in a 200 $\mu$m deep capillary tube 
%The I-N phase transition region is inversely proportional to the average filament length.

%, which become visible in the order of minutes after polymerization initiation

\begin{figure}[b]
\includegraphics[width=8.5cm]{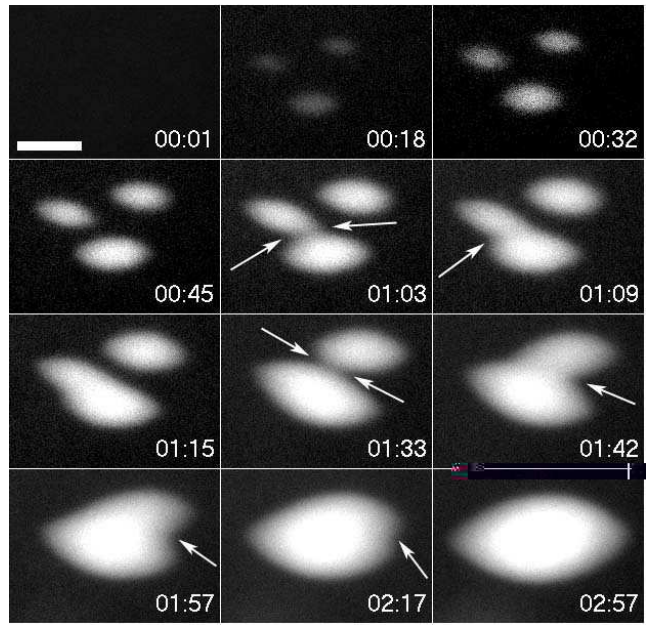}
\caption{\label{fig4} Growth of three N tactoids and their sequential coalescence to form one final tactoid, recorded under a polarization microscope. The arrows indicate  surface distortions while the tactoids are fusing. Time format is  hour:minutes. Scale bar is 20 $\mu m$}
\end{figure}

When actin is polymerized at slightly above or below the co-existing I or N concentrations, nascent tactoids are nucleated in the transition region for F-actin with $\ell \leq 2 \mu m$. Coalescence of tactoidal droplets is an efficient form of growth. Fig.~\ref{fig4} shows how N tactoids coalesce, viewed under a polarization microscope. Initially the tactoids are separated from each other. Upon growth, two tactoids reach close proximity and fuse. Once two tactoids have coalesced the final shape is again a similar tactoid. The process repeats itself as the 3rd tactoid coalesces to form one final tactoid. One charactistic of such a large molecular system is its slow dynamics, as the sequence in display took nearly 3 hours.

\begin{figure}[t]
\includegraphics[width=8.0cm]{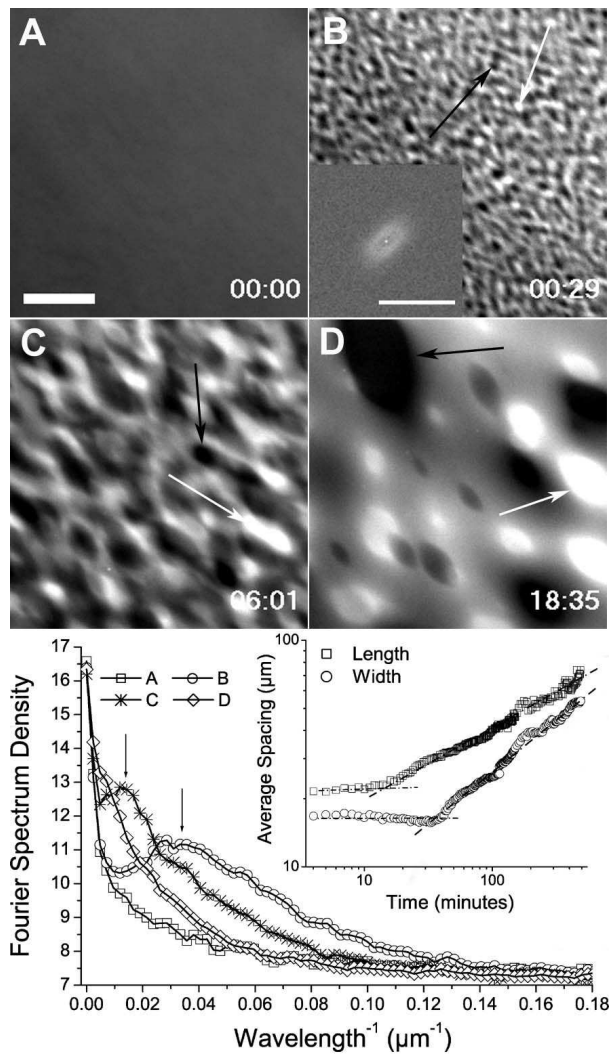}
\caption{\label{fig3}Tactoid nucleation and growth for 10.8 mg/ml actin concentration and $\ell = 1 \mu m$, recorded by polarization microscopy. (\textbf{A}) Uniform sample immediately after polymerization initiation. The scale bar represents 100 $\mu m$. (\textbf{B}) Appearance of tactoidal droplets. Inset: 2D-FFT of the pattern, showing a range of sizes and orientations; bar represents a wavelength of $ 2 \mu m$. (\textbf{C,D}) Late stage growth of N and I tactoids. White arrows point to N tactoids and black ones to I tactoids. Time stamp format is hours:minutes.  The plot shows peak (arrows) progression in the Fourier spectrum density, which corresponds to the dominant size of granules. Leggend is corresponding to the images letter designation. Insert, characteristic length and width as functions of time. The dashed lines through the late stage data represent power-law fits, with exponents of 0.31 and 0.47 for the growth of length and width, respectively.}
\end{figure}

When the actin concentration is close to midway between the co-existing I and N concentrations (both are sensitive functions of $\ell $), we reproducibly observed the phenomenon of spinodal decomposition. At the high concentration of several mg/ml, F-actin polymerization occurs within seconds following the addition of KCl and MgCl$_2$ \cite{Carlier1984}. Therefore, the solution becomes weakly aligned by the shear flow as it is injected into the capillary tube. Fig.~\ref{fig3}\textbf{A}-\textbf{D} show a time sequence of tactoidal growth from initiation of actin polymerization to the formation of large tactoids. Fig. \ref{fig3}\textbf{A} represents the actin solution immediately after it was prepared for observation, which took about 30 seconds. A granular structure appears throughout the capillary tube within minutes after initiation of polymerization. The characteristic  domain size is determined by 2D-FFT and is $\sim 17 \mu m\times\sim 22 \mu m$ (see inset on the bottom plot). The domain size appears to be nearly constant within the first 15 or 30 min for length and width. The dominant size of the spinodal growth can be explained by adding a density gradient term to the free energy of the system following the classical Cahn-Hilliard treatment \cite{Jones}. Based on the observed peak wave factor $q_{m}=1/\lambda$, we predict that the thickness of the I-N interface $\xi \sim \lambda /4\pi = 1.4 \mu m$ \cite{Larson}, assuming that filaments tend to align parallel to the interface.  By the 30 min time point, both I and N droplets are discernable (Fig.\ref{fig3}\textbf{B}). The late stage growth lasts for many hours. For instance, Fig.\ref{fig3}\textbf{C} represents the time point when coalescence has become the main form of tactoidal growth. The large tactoids in Fig.~\ref{fig3}\textbf{D} are formed by both continuous growth and occasional coalescence of smaller tactoids. The plot in Fig \ref{fig3} shows progression of the radial peak in the Fourier spectrum density, whereas the tactoidal length and width shown in the inset were obtained by analyzing line plots along the long and short axes separately.  

It is important to note that  the actin tactoids we report here are fundamentally different from what we recently reported \cite{TangHyeran2005}. Actin granules reported in our recent study were induced by an actin crosslinking protein, alpha-actinin. As a result, the density of actin in the published actin granules is over 10 times higher than in this study. The dense actin granules observed in our recent report are clearly discernable with phase contrast imaging, which is not the case for either I or N droplets reported here. Also, in the present study, we have not observed variant shapes such as triangular tactoids. The tactoids formed due to I-N separation easily disappear upon dilution, unlike the permenant tactoids due to crosslinking of alpha-actinin. Nevertheless, it is remarkable that in both cases the much similar tactoidal shape prodominates. We argue that the tactoidal shape is general among granules consisting of long and stiff filaments, which is dictated by the minimization of the surface energy while accomodating the long and stiff constituent filaments. It is thus not surprising that tactoids of similar shapes are found in two distinct types of actin granules, and in concentrated suspension of Tobacco Mosaic Virus (TMV) \cite{TMV} and filamentous phage fd \cite{DogicPRL2004}, as well.  

In conclusion, we have shown new features of I-N phase transition of F-actin solutions as a function of $\ell$. At $\ell \geq 3 \mu m$, the I-N phase transition is continuous, consistent with previous findings \cite{Viamontes2003,Coppin}. However, biphasic behavior characteristic of a first order transition is observed for $\ell \leq 2 \mu m$, including both phenomena of nucleation-growth and spinodal decomposition. Tactoidal droplets of either I or N domains form in the N or I background. Tactoids of both phases can also co-exist with a weakly aligned background state, suggesting slow kinetics and metastability. The process for tactoidal growth involves both constant recruitment of the surrounding filaments and coalescence of existing neighboring tactoids.

This work is supported by the National Science Foundation (NSF DMR 0405156) and the Petroleum Research Fund, administered by the American Chemical Society. We thank Professors Robert Meyer, Robert Pelcovits, Tom Powers and Jim Valles for valuable suggestions.

\bibliography{Bib}

\begin{thebibliography}{29}
\expandafter\ifx\csname natexlab\endcsname\relax\def\natexlab#1{#1}\fi
\expandafter\ifx\csname bibnamefont\endcsname\relax
  \def\bibnamefont#1{#1}\fi
\expandafter\ifx\csname bibfnamefont\endcsname\relax
  \def\bibfnamefont#1{#1}\fi
\expandafter\ifx\csname citenamefont\endcsname\relax
  \def\citenamefont#1{#1}\fi
\expandafter\ifx\csname url\endcsname\relax
  \def\url#1{\texttt{#1}}\fi
\expandafter\ifx\csname urlprefix\endcsname\relax\def\urlprefix{URL }\fi
\providecommand{\bibinfo}[2]{#2}
\providecommand{\eprint}[2][]{\url{#2}}

\bibitem[{\citenamefont{Alberts and
  et~al.}(2002)}]{Molecular_biology_of_the_cell}
\bibinfo{author}{\bibfnamefont{B.}~\bibnamefont{Alberts}} \bibnamefont{and}
  \bibinfo{author}{\bibnamefont{et~al.}}, \emph{\bibinfo{title}{Mol. biol. of
  the cell}} (\bibinfo{publisher}{Garland}, \bibinfo{address}{New York},
  \bibinfo{year}{2002}), \bibinfo{edition}{4th} ed.

\bibitem[{\citenamefont{Holmes et~al.}(1990)\citenamefont{Holmes, Popp,
  Gebhard, and Kabsch}}]{HolmesPoppGebhardKabsch}
\bibinfo{author}{\bibfnamefont{K.~C.} \bibnamefont{Holmes}},
  \bibinfo{author}{\bibfnamefont{D.}~\bibnamefont{Popp}},
  \bibinfo{author}{\bibfnamefont{W.}~\bibnamefont{Gebhard}}, \bibnamefont{and}
  \bibinfo{author}{\bibfnamefont{W.}~\bibnamefont{Kabsch}},
  \bibinfo{journal}{Nature} \textbf{\bibinfo{volume}{347}}, \bibinfo{pages}{44}
  (\bibinfo{year}{1990}).

\bibitem[{\citenamefont{Sept et~al.}(1999)\citenamefont{Sept, Xu, Pollard, and
  McCammon}}]{Sept_Xu_Pollard_McCammon}
\bibinfo{author}{\bibfnamefont{D.}~\bibnamefont{Sept}},
  \bibinfo{author}{\bibfnamefont{J.}~\bibnamefont{Xu}},
  \bibinfo{author}{\bibfnamefont{T.~D.} \bibnamefont{Pollard}},
  \bibnamefont{and} \bibinfo{author}{\bibfnamefont{J.~A.}
  \bibnamefont{McCammon}}, \bibinfo{journal}{Biophys. J.}
  \textbf{\bibinfo{volume}{77}}, \bibinfo{pages}{2911} (\bibinfo{year}{1999}).

\bibitem[{\citenamefont{Isambert et~al.}(1995)\citenamefont{Isambert, Vernier,
  Maggs, and et~al.}}]{Isambert_Vernier_Maggs_Carlier}
\bibinfo{author}{\bibfnamefont{H.}~\bibnamefont{Isambert}},
  \bibinfo{author}{\bibfnamefont{P.}~\bibnamefont{Vernier}},
  \bibinfo{author}{\bibfnamefont{A.~C.} \bibnamefont{Maggs}}, \bibnamefont{and}
  \bibinfo{author}{\bibnamefont{et~al.}}, \bibinfo{journal}{J. of Biol. Chem.}
  \textbf{\bibinfo{volume}{270}}, \bibinfo{pages}{11437}
  (\bibinfo{year}{1995}).

\bibitem[{\citenamefont{Gittes and
  et~al.}(1993)}]{Gittes_Mickey_Nettlenton_Howard}
\bibinfo{author}{\bibfnamefont{F.}~\bibnamefont{Gittes}} \bibnamefont{and}
  \bibinfo{author}{\bibnamefont{et~al.}}, \bibinfo{journal}{J. of cell biol.}
  \textbf{\bibinfo{volume}{120}}, \bibinfo{pages}{923} (\bibinfo{year}{1993}).

\bibitem[{\citenamefont{Fujiwara et~al.}(2002)\citenamefont{Fujiwara,
  Takahashi, Tadakuma, Funatsu, and Ishiwata}}]{Fujiwara2002}
\bibinfo{author}{\bibfnamefont{I.}~\bibnamefont{Fujiwara}},
  \bibinfo{author}{\bibfnamefont{S.}~\bibnamefont{Takahashi}},
  \bibinfo{author}{\bibfnamefont{H.}~\bibnamefont{Tadakuma}},
  \bibinfo{author}{\bibfnamefont{T.}~\bibnamefont{Funatsu}}, \bibnamefont{and}
  \bibinfo{author}{\bibfnamefont{S.}~\bibnamefont{Ishiwata}},
  \bibinfo{journal}{Nature Cell Biol.} \textbf{\bibinfo{volume}{4}},
  \bibinfo{pages}{666} (\bibinfo{year}{2002}).

\bibitem[{\citenamefont{Coppin and Leavis}(1992)}]{Coppin}
\bibinfo{author}{\bibfnamefont{C.}~\bibnamefont{Coppin}} \bibnamefont{and}
  \bibinfo{author}{\bibfnamefont{P.}~\bibnamefont{Leavis}},
  \bibinfo{journal}{Biophys. J.} \textbf{\bibinfo{volume}{63}},
  \bibinfo{pages}{794} (\bibinfo{year}{1992}).

\bibitem[{\citenamefont{Furakawa and et~al.}(1993)}]{Furakawa}
\bibinfo{author}{\bibfnamefont{R.}~\bibnamefont{Furakawa}} \bibnamefont{and}
  \bibinfo{author}{\bibnamefont{et~al.}}, \bibinfo{journal}{Biochemistry}
  \textbf{\bibinfo{volume}{32}}, \bibinfo{pages}{12346} (\bibinfo{year}{1993}).

\bibitem[{\citenamefont{Viamontes and Tang}(2003)}]{Viamontes2003}
\bibinfo{author}{\bibfnamefont{J.}~\bibnamefont{Viamontes}} \bibnamefont{and}
  \bibinfo{author}{\bibfnamefont{J.~X.} \bibnamefont{Tang}},
  \bibinfo{journal}{Phys. Rev. E} \textbf{\bibinfo{volume}{67}},
  \bibinfo{pages}{040701(R)} (\bibinfo{year}{2003}).

\bibitem[{\citenamefont{Gardel et~al.}(2003)\citenamefont{Gardel, Valentine,
  Crocker, Bausch, and Weitz}}]{GardelPRL2003}
\bibinfo{author}{\bibfnamefont{M.~L.} \bibnamefont{Gardel}},
  \bibinfo{author}{\bibfnamefont{M.~T.} \bibnamefont{Valentine}},
  \bibinfo{author}{\bibfnamefont{J.~C.} \bibnamefont{Crocker}},
  \bibinfo{author}{\bibnamefont{Bausch}}, \bibnamefont{and}
  \bibinfo{author}{\bibfnamefont{D.~A.} \bibnamefont{Weitz}},
  \bibinfo{journal}{Phys Rev. Lett.} \textbf{\bibinfo{volume}{91}},
  \bibinfo{pages}{158302} (\bibinfo{year}{2003}).

\bibitem[{\citenamefont{Suzuki and et~al.}(1991)}]{Suzuki}
\bibinfo{author}{\bibfnamefont{A.}~\bibnamefont{Suzuki}} \bibnamefont{and}
  \bibinfo{author}{\bibfnamefont{T.~M.} \bibnamefont{et~al.}},
  \bibinfo{journal}{Biophys. J.} \textbf{\bibinfo{volume}{59}},
  \bibinfo{pages}{25} (\bibinfo{year}{1991}).

\bibitem[{\citenamefont{Onsager}(1949)}]{Onsager}
\bibinfo{author}{\bibfnamefont{L.}~\bibnamefont{Onsager}},
  \bibinfo{journal}{Ann. NY. Acad. Sci.} \textbf{\bibinfo{volume}{51}},
  \bibinfo{pages}{627} (\bibinfo{year}{1949}).

\bibitem[{\citenamefont{Flory}(1969)}]{Flory}
\bibinfo{author}{\bibfnamefont{P.~J.} \bibnamefont{Flory}},
  \emph{\bibinfo{title}{Statistical Mechanics of Chain Molecules}}
  (\bibinfo{publisher}{Interscience Publishers}, \bibinfo{address}{New York},
  \bibinfo{year}{1969}).

\bibitem[{\citenamefont{Lammert et~al.}(1993)\citenamefont{Lammert, Rokhar, and
  Toner}}]{LRTPRL}
\bibinfo{author}{\bibfnamefont{P.}~\bibnamefont{Lammert}},
  \bibinfo{author}{\bibfnamefont{D.}~\bibnamefont{Rokhar}}, \bibnamefont{and}
  \bibinfo{author}{\bibfnamefont{J.}~\bibnamefont{Toner}},
  \bibinfo{journal}{Phys. Rev. Lett.} \textbf{\bibinfo{volume}{70}},
  \bibinfo{pages}{1650} (\bibinfo{year}{1993}).

\bibitem[{\citenamefont{Lammert et~al.}(1995)\citenamefont{Lammert, Rokhar, and
  Toner}}]{LRTPRE}
\bibinfo{author}{\bibfnamefont{P.}~\bibnamefont{Lammert}},
  \bibinfo{author}{\bibfnamefont{D.}~\bibnamefont{Rokhar}}, \bibnamefont{and}
  \bibinfo{author}{\bibfnamefont{J.}~\bibnamefont{Toner}},
  \bibinfo{journal}{Phys. Rev. E} \textbf{\bibinfo{volume}{52}},
  \bibinfo{pages}{1778} (\bibinfo{year}{1995}).

\bibitem[{\citenamefont{Prinsen and van~der
  Schoot}(2003)}]{Prinsen_van_der_Schoot}
\bibinfo{author}{\bibfnamefont{P.}~\bibnamefont{Prinsen}} \bibnamefont{and}
  \bibinfo{author}{\bibfnamefont{P.}~\bibnamefont{van~der Schoot}},
  \bibinfo{journal}{Phys. Rev. E} \textbf{\bibinfo{volume}{68}},
  \bibinfo{pages}{021701(1)} (\bibinfo{year}{2003}).

\bibitem[{\citenamefont{Drzaic}(1995)}]{Drzaic}
\bibinfo{author}{\bibfnamefont{P.}~\bibnamefont{Drzaic}},
  \emph{\bibinfo{title}{Liquid Crystal Dispersions}} (\bibinfo{publisher}{World
  Scientific}, \bibinfo{address}{Singapore}, \bibinfo{year}{1995}).

\bibitem[{\citenamefont{Pardee and Spudich}(1982)}]{Pardee}
\bibinfo{author}{\bibfnamefont{J.}~\bibnamefont{Pardee}} \bibnamefont{and}
  \bibinfo{author}{\bibfnamefont{J.}~\bibnamefont{Spudich}},
  \bibinfo{journal}{Methods Cell Biol} \textbf{\bibinfo{volume}{24}},
  \bibinfo{pages}{271} (\bibinfo{year}{1982}).

\bibitem[{\citenamefont{Janmey et~al.}(1986)\citenamefont{Janmey, Peetermans,
  Zaner, and et~al.}}]{JanmeyPeetermans}
\bibinfo{author}{\bibfnamefont{P.}~\bibnamefont{Janmey}},
  \bibinfo{author}{\bibfnamefont{J.}~\bibnamefont{Peetermans}},
  \bibinfo{author}{\bibfnamefont{K.}~\bibnamefont{Zaner}}, \bibnamefont{and}
  \bibinfo{author}{\bibnamefont{et~al.}}, \bibinfo{journal}{J. Biol. Chem}
  \textbf{\bibinfo{volume}{261}}, \bibinfo{pages}{8357} (\bibinfo{year}{1986}).

\bibitem[{\citenamefont{Tang and Janmey}(1996)}]{TangJanmey}
\bibinfo{author}{\bibfnamefont{J.~X.} \bibnamefont{Tang}} \bibnamefont{and}
  \bibinfo{author}{\bibfnamefont{P.~A.} \bibnamefont{Janmey}},
  \bibinfo{journal}{J. Biol. Chem.} \textbf{\bibinfo{volume}{271}},
  \bibinfo{pages}{8556} (\bibinfo{year}{1996}).

\bibitem[{\citenamefont{Oldenbourg and Mei}(1995)}]{OldenbourgMei}
\bibinfo{author}{\bibfnamefont{R.}~\bibnamefont{Oldenbourg}} \bibnamefont{and}
  \bibinfo{author}{\bibfnamefont{G.}~\bibnamefont{Mei}}, \bibinfo{journal}{J.
  of Micro.} \textbf{\bibinfo{volume}{180}}, \bibinfo{pages}{140}
  (\bibinfo{year}{1995}).

\bibitem[{\citenamefont{{Shribak} and {Oldenbourg}}(2003)}]{ShribakOldenbourg}
\bibinfo{author}{\bibfnamefont{M.}~\bibnamefont{{Shribak}}} \bibnamefont{and}
  \bibinfo{author}{\bibfnamefont{R.}~\bibnamefont{{Oldenbourg}}},
  \bibinfo{journal}{\ao} \textbf{\bibinfo{volume}{42}}, \bibinfo{pages}{3009}
  (\bibinfo{year}{2003}).

\bibitem[{\citenamefont{Bees and Hill}(1997)}]{Bees_and_Hill}
\bibinfo{author}{\bibfnamefont{M.}~\bibnamefont{Bees}} \bibnamefont{and}
  \bibinfo{author}{\bibfnamefont{N.}~\bibnamefont{Hill}}, \bibinfo{journal}{J.
  of Exp. Biol.} \textbf{\bibinfo{volume}{200}}, \bibinfo{pages}{1515}
  (\bibinfo{year}{1997}).

\bibitem[{\citenamefont{Carlier et~al.}(1984)\citenamefont{Carlier, Pantaloni,
  and Korn}}]{Carlier1984}
\bibinfo{author}{\bibfnamefont{M.-F.} \bibnamefont{Carlier}},
  \bibinfo{author}{\bibfnamefont{D.}~\bibnamefont{Pantaloni}},
  \bibnamefont{and} \bibinfo{author}{\bibfnamefont{E.~D.} \bibnamefont{Korn}},
  \bibinfo{journal}{J. of Biol. Chem.} \textbf{\bibinfo{volume}{259}},
  \bibinfo{pages}{9983} (\bibinfo{year}{1984}).

\bibitem[{\citenamefont{Jones}(2002)}]{Jones}
\bibinfo{author}{\bibfnamefont{R.~A.~L.} \bibnamefont{Jones}},
  \emph{\bibinfo{title}{Soft Condensed Matter}} (\bibinfo{publisher}{Oxford
  University Press}, \bibinfo{address}{New York}, \bibinfo{year}{2002}).

\bibitem[{\citenamefont{Larson}(1999)}]{Larson}
\bibinfo{author}{\bibfnamefont{R.~G.} \bibnamefont{Larson}},
  \emph{\bibinfo{title}{The Structure and Rheology of Complex Fluids}}
  (\bibinfo{publisher}{Oxford University Press}, \bibinfo{address}{New York},
  \bibinfo{year}{1999}).

\bibitem[{\citenamefont{Tang et~al.}(2005)\citenamefont{Tang, Kang, and
  Jia}}]{TangHyeran2005}
\bibinfo{author}{\bibfnamefont{J.~X.} \bibnamefont{Tang}},
  \bibinfo{author}{\bibfnamefont{H.}~\bibnamefont{Kang}}, \bibnamefont{and}
  \bibinfo{author}{\bibfnamefont{J.}~\bibnamefont{Jia}},
  \bibinfo{journal}{Langmuir} \textbf{\bibinfo{volume}{21}},
  \bibinfo{pages}{2789} (\bibinfo{year}{2005}).

\bibitem[{\citenamefont{Freundlich}(1938)}]{TMV}
\bibinfo{author}{\bibfnamefont{H.}~\bibnamefont{Freundlich}},
  \bibinfo{journal}{J. Phys. Chem.} \textbf{\bibinfo{volume}{41}},
  \bibinfo{pages}{1151} (\bibinfo{year}{1938}).

\bibitem[{\citenamefont{Dogic and et~al.}(2004)}]{DogicPRL2004}
\bibinfo{author}{\bibfnamefont{Z.}~\bibnamefont{Dogic}} \bibnamefont{and}
  \bibinfo{author}{\bibnamefont{et~al.}}, \bibinfo{journal}{Phys. Rev. Lett.}
  \textbf{\bibinfo{volume}{92}}, \bibinfo{pages}{125503}
  (\bibinfo{year}{2004}).

\end{thebibliography}

\end{document}